\documentclass[twocolumn,superscriptaddress,secnumarabic,
amssymb,amsmath,nobibnotes,aps,prd,showkeys,showpacs,nofootinbib]{revtex4}%
\usepackage{graphicx}
\usepackage{epsf}
\usepackage{bm}
\usepackage{amsmath}
\usepackage{amsfonts}
\usepackage{amssymb}
\usepackage{epstopdf}
\usepackage{natbib}
\usepackage{color}
\providecommand{\U}[1]{\protect\rule{.1in}{.1in}}

\newcommand{\be}{\begin{equation}}
\newcommand{\ee}{\end{equation}}

\newcommand{\mincir}{\raise
-3.truept\hbox{\rlap{\hbox{$\sim$}}\raise4.truept\hbox{$<$}\ }}
\newcommand{\magcir}{\raise
-3.truept\hbox{\rlap{\hbox{$\sim$}}\raise4.truept\hbox{$>$}\ }}

\newtheorem{remark}{Remark}[section]

\begin{document}

\title{Gravitational production of dark matter in the  Peebles-Vilenkin model}

\author{Jaume   Haro}
\email{jaime.haro@upc.edu}
\affiliation{Departament de Matem\`atiques, Universitat Polit\`ecnica de Catalunya, Diagonal 647, 08028 Barcelona, Spain}



\thispagestyle{empty}

\begin{abstract}
The gravitational production of superheavy dark matter, in the Peebles-Vilenkin quintessential inflation model,  is studied in two different scenarios: When the particles, whose decay products  reheat the universe after the end of the inflationary period, 
are created gravitationally, and when are produced via {\it instant preheating}. We show that the viability of both scenarios requires that the mass of the superheavy  dark matter to be 
approximately  {between $10^{16}$ and  $10^{17}$ GeV.}
\end{abstract}

\vspace{0.5cm}

\maketitle
\section{Introduction}

Quintessential inflation, which was addressed for the first time by
 Peebles and Vilenkin (PV) in \cite{pv}, is an attempt to unify  inflation and quintessence via a single scalar field whose potential
allows inflation while at late time provides quintessence (see for instance \cite{dimopoulos}).
A remarkable property of the  PV model  is that it contains an abrupt phase transition from inflation to  kination 
(a regime where all the energy density of the inflation turns into kinetic), where   the adiabatic regime is broken and, thus, particles could be gravitationally created \cite{ford, Damour}. This leads to the possibility to explain the abundance of dark matter through the gravitational production of superheavy particles \cite{hashiba, hashiba1}, although gravitational production of dark matter could also occur in standard inflation during the oscillations of the inflaton field \cite{kolb1, ema,kolb2} {{}(see also the early papers
\cite{Starobinsky, Starobinsky1, Vilenkin}).}

\

Considering the gravitational production of  two kinds of superheavy particles: $X$-particles, conformally coupled with gravity,
whose energy density after their decay and later thermalization of decay products will dominate the energy density of the scalar field in order to match with the Hot Big Bang (HBB), and dark $Y$-particles  which  are only gravitationally interacting massive particles (GIMP), 
we will show that  the PV model preserves the Big Bang Nucleosynthesis (BBN) success. More precisely, the overproduction of Gravitational Waves (GWs) does not disturb the BBN   for $X$-particles and $Y$-particles with masses in the range of {{} $10^{15}-10^{16}$  GeV  and $10^{16}-10^{17}$ GeV} respectively,  leading to a maximum reheating temperature in the {{} GeV} regime.

\

{On the contrary, for massless conformally coupled $X$-particles  produced via {\it instant preheating} (see \cite{fkl0} for a detailed discussion of this mechanism of particle creation),
 assuming that  $Y$-GIMP, {{} which are gravitationally produced}, are the constituent of the dark matter, 
the viability of the model requires that the mass of the superheavy $Y$-particles is approximately
of the order of {{} $10^{16}$ GeV}, yielding a reheating temperature around $10^8$ GeV.

}

\

{{} The work is structured as follows: In Section $2$ and improved version of the well-known Peebles-Vilenkin model for quintessential inflation is presented. Section $3$ is devoted to the study of gravitational production of  superheavy $X$-particles whose decaying products reheat
the universe and  superheavy  $Y$-particles which are the responsible for the abundance of dark matter. In addition, we show  how to overpass the constrains coming from the overproduction of gravitational waves. In Section $4$ we consider the case in which the $X$-particles are produced via instant preheating. The  dynamics of the scalar field, for the improved model proposed in Section $2$,  is studied in detail in Section $5$, and finally, we present the conclusions of our study in Section $6$.

}

\section{The Peebles-Vilenkin model}
It is well-known that in quintessential inflation the number of e-folds from the pivot scale exiting the Hubble radius  to the end of inflation is greater than $60$. 
For this reason,
in order that  the theoretical values of the spectral index and the ratio of tensor to scalar perturbations enters  in their marginalized joint confidence contour
in the plane
$(n_s,r)$ at $2\sigma$ C.L., 
we have changed the quartic inflationary potential of the original  
 PV quintessential inflation model with a quadratic one, obtaining:
\begin{eqnarray}\label{pv}
V(\varphi)=\left\{\begin{array}{ccc}
\frac{1}{2}m^2(\varphi^2+M^2)& \mbox{for}& \varphi\leq 0\\
\frac{1}{2}m^2\frac{M^6}{\varphi^4+M^4} &\mbox{for}& \varphi\geq 0,
\end{array}\right.
\end{eqnarray}
where $m$ is the mass of the scalar field and $M\sim 10$ GeV, is an small mass that has to be calculated numerically \cite{hap18}.

\

{{} As we can see in the Figure $1$,
the spectral index and the tensor/scalar ratio 
enter perfectly in the  two dimensional marginalized joint confidence contour at $2\sigma$ Confidence Level (CL) for the Planck TT, TE, EE + low E 
and for the Planck TT, TE, EE + low E + lensing  likelihood \cite{planck18}. In addition, if one  wants that the model enters, at $2\sigma$ CL,
in the contour for the Planck TT, TE, EE + low E + lensing +BK14+BAO likelihood, i.e., taking into account gravitational waves, one has to replace the inflationary piece
of the potential by a plateau-like potential \cite{plateau} or $\alpha$-attractors \cite{attractor0,attractor1,attractor2} such as an Starobinsky-type potential \cite{riotto}
{{}\begin{eqnarray}\label{PVimproved}
V(\varphi)=\left\{\begin{array}{ccc}
\lambda M_{pl}^4\left(1-e^{\sqrt{\frac{2}{3}}\frac{\varphi}{M_{pl}}}\right)^2 + \lambda \tilde{M}^4 & \mbox{for} & \varphi\leq 0\\
\lambda\frac{\tilde{M}^8}{\varphi^{4}+\tilde{M}^4} &\mbox{for} & \varphi\geq 0,\end{array}
\right.
\end{eqnarray}}
where $\lambda$ is a dimensionless parameter of the order $10^{-10}$, {{} and now $\tilde{M}\sim 10^5$ GeV (see \cite{pv})}. Effectively, for the potential (\ref{PVimproved}) one has  (see for instance \cite{attractor2})
\begin{eqnarray}
n_s\cong 1-\frac{2}{N}, \quad  \mbox{ and }  \quad r\cong\frac{12}{N^2},
\end{eqnarray}
where $N$ is the number of e-folds. Thus,  as we have already explained, since  in quintessential inflation the number of e-folds is greater than $60$ one gets that 
$r< 0.0034$, and clearly, the spectral index and the tensor/scalar ratio enters  at $2\sigma$ CL,
in the contour for the Planck TT, TE, EE + low E + lensing +BK14+BAO likelihood (see Figure $1$).

 \begin{figure}
\includegraphics[width=0.54\textwidth]{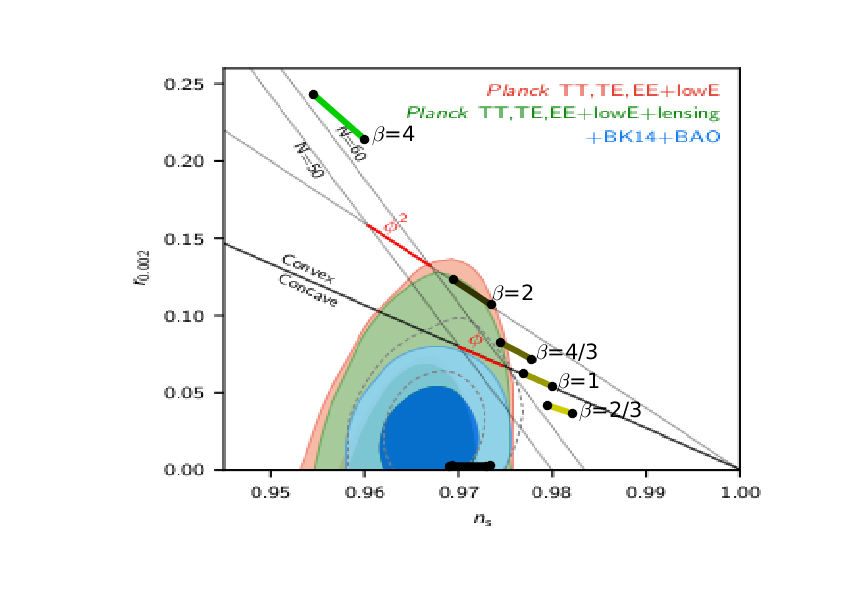}
{{}\caption{Marginalized joint confidence contours for $(n_s,r)$ at $1\sigma$ and $2\sigma$ Confidence Level (CL).  Considering the inflationary piece of the potential as 
$V=\lambda \phi^{\beta}$, in
quintessential inflation, for the values of $\beta=4,2, 3/4, 1, 2/3$, we have
drawn the curves  from  $65$ to $75$ e-folds (see the green, which correspond to the original P-V model, and  black curves). And  when one considers the standard inflation, for $\beta=2, 1$, the curves have been drawn in red from $50$ to $60$ $e$-folds.
As one can see, the quadratic potential ($V \propto \phi^2$), which is disregarded in standard inflation at greater than $2\sigma$ CL
from a combination of Planck and BICEP2 limits on the tensor-to-scalar 
ratio \cite{planck18}, is favored  for some likelihoods in quintessential inflation. In the lower part of the image there is the curve  for the potential \eqref{PVimproved}. The value of $r$ is nearly $0$ and, if considering all Planck likelihoods, it stands within the $2\sigma$ CL for $65\lesssim N\lesssim 75$.}}
\label{fig:n-r}
\end{figure}

\

{}
\begin{remark}
The first piece of the potential (\ref{PVimproved})  is obtained when one deals with $R^2$ gravity in the Einstein Frame \cite{riotto}, and the tail, which is the same used
in [1],  comes from SUSY QED \cite{SUSY}.\end{remark}

\begin{remark}
The second derivative of the potentials (\ref{pv}) and (\ref{PVimproved}) has a jump discontinuity  at the beginning of kination, but its physical origin is not discussed in
the present work. However, one may argue, as was shown in \cite{starobinsky} where the discontinuity of the second derivative of the potential appears during inflation, that its origin
could be due to a second-order phase transition of another scalar field coupled with the  field $\varphi$. This is a point that deserves future investigation.

\end{remark}

{{}

\

To calculate $H_{kin}$, the value of the Hubble parameter at the beginning of kination 
 for the model (\ref{PVimproved}),
first of all we calculate the slow roll parameters:
Denoting by  $\epsilon_*=\frac{M_{pl}^2}{2}\left(\frac{V_{\varphi}(\varphi_*)}{V(\varphi_*)}  \right)^2$ and 
$\eta_* ={M_{pl}^2}\frac{V_{\varphi\varphi}(\varphi_*)}{V(\varphi_*)}$ the values of the slow roll parameters  and by $\varphi_*$ the value of the scalar field  when the pivot scale exits the Hubble radius,  since the mass {{} $\tilde{M}$ satisfies $\tilde{M}\ll M_{pl}$}, 
 one has $
\epsilon_*\cong \frac{4}{3}e^{2\sqrt{\frac{2}{3}}\frac{\varphi_*}{M_{pl}}}  $
$\eta_*=    -\frac{4}{3}e^{\sqrt{\frac{2}{3}}\frac{\varphi_*}{M_{pl}}} ,$
and thus, the spectral index is given by \cite{btw}
\begin{eqnarray}
1-n_s\cong 6\epsilon_*-2\eta_*\cong \frac{8}{3}e^{\sqrt{\frac{2}{3}}\frac{\varphi_*}{M_{pl}}},
\end{eqnarray}
meaning that
\begin{eqnarray}
 \varphi_*\cong \sqrt{\frac{3}{2}}M_{pl}\ln\left(\frac{3}{8} (1-n_s)\right).
\end{eqnarray}

On the other hand, the observational estimation of the power spectrum of the scalar perturbations when the pivot scale leaves the Hubble radius is ${\mathcal P}_{\zeta}\cong \frac{H_*^2}{8\pi^2M_{pl}^2\epsilon_*}\sim 2\times 10^{-9}$ \cite{btw}. 
Since during the slow roll regime the kinetic energy density is negligible compared with the potential one, we will have 
$H_*^2\cong \frac{\lambda}{3} M_{pl}^2$, and using the relation $\epsilon_*=\frac{3}{16}(1-n_s)^2$ one gets
\begin{eqnarray} 
\lambda\sim 9\pi^2(1-n_s)^2\times 10^{-9}.
\end{eqnarray}

Taking into account that the observational value of the spectral index is $n_s=0.968\pm 0.006$ \cite{Planck}, if one chooses  its central value one gets
\begin{eqnarray} \lambda=9\times 10^{-11} \quad  \mbox{and} \quad
\varphi_*\cong -5.42 M_{pl}. 
\end{eqnarray}

\

Now, taking into account that  inflation ends when $\epsilon=1$, that is, when 
 $e^{\sqrt{\frac{2}{3}}\frac{\varphi}{M_{pl}}}=\sqrt{3}(2-\sqrt{3}) $. 
This means that the value of the potential energy at the end of inflation is approximately
$\frac{3}{2}\lambda (4-2\sqrt{3})^2 M_{pl}^4$ which is many orders greater than $V(0)=\lambda \tilde{M}^4$, because $\tilde{M}\ll M_{pl}$. Thus, we can safely conclude that the kination phase, i.e. when practically all energy density is kinetic,   {{} has already started when the field  $\varphi$ crosses the origin}
   (see also the Figure $2$, where one can see that the maximum of the velocity of the scalar field is obtained very close to $\varphi=0$.). Then, {{} to simplify, we can consider that kination starts at $\varphi_{kin}=0$}, and   to obtain the value of the Hubble rate at the beginning of kination, namely $H_{kin}$,
 we have to solve numerically the conservation equation 
\begin{eqnarray}\label{KG}
\ddot{\varphi}+3
\sqrt{\frac{\frac{\dot{\varphi}^2}{2}+V(\varphi)}{3M_{pl}^2}}\dot{\varphi}+V_{\varphi}=0,
\end{eqnarray}
with initial conditions $\varphi_*=-5.42 M_{pl}$ and $\dot{\varphi}_*=0$ (obviously, one can choose other similar initial conditions and the result has to be practically the same, because the inflationary dynamics is an attractor).

\begin{figure}
\includegraphics[width=0.5\textwidth]{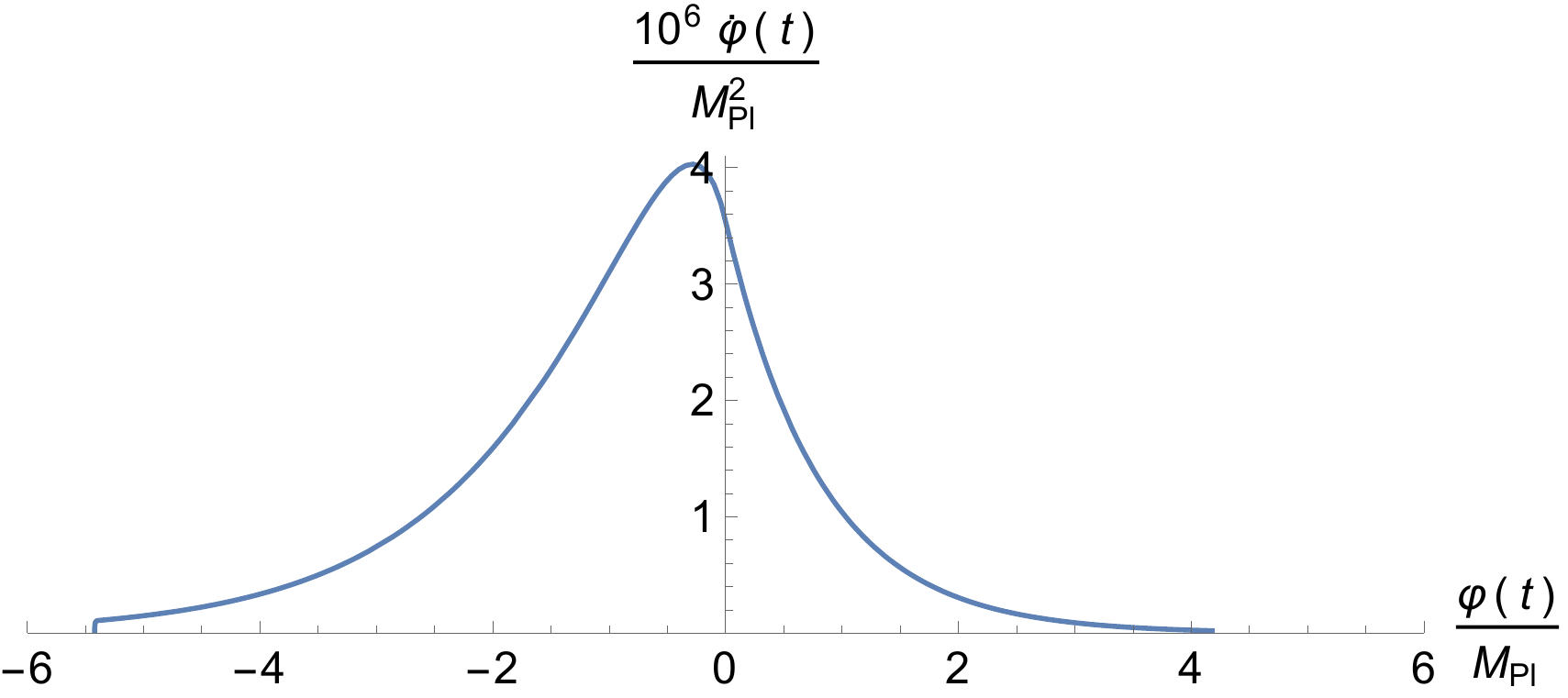}
{{}\caption{Evolution of the velocity of the scalar field obtained integrating the equation (\ref{KG})  with initial conditions when the pivot scale leaves the Hubble horizon, i.e., for
$\varphi_*=-5.42 M_{pl}$ and $\dot{\varphi}_*=0$. }}
\end{figure}

\

 Using event-driven integration with an ode RK78 integrator, when $\varphi$ vanishes  one gets
$\dot{\varphi}_{kin}=3.54\times 10^{-6} M_{pl}^2$, and thus
 \begin{eqnarray}
  H_{kin}=\frac{\dot{\varphi}_{kin}}{\sqrt{6} M_{pl}}\cong 1.44\times 10^{-6} M_{pl},
  \end{eqnarray}
and 
\begin{eqnarray}\label{8}
\rho_{\varphi, kin}\cong  6.26\times 10^{-12}M_{pl}^4.
\end{eqnarray}

}

}

\

Coming back to the PV model (\ref{pv}),
one has $
\epsilon_*=\eta_*=\frac{2M_{pl}^2}{\varphi^2_*},$
and thus, the spectral index is given by 
\begin{eqnarray}
1-n_s=6\epsilon_*-2\eta_*=\frac{8M_{pl}^2}{\varphi^2_*}\nonumber \\ \Longrightarrow \varphi_*=-\sqrt{\frac{8}{1-n_s}}M_{pl},
\end{eqnarray} 
{{}
and using the formula of the power spectrum of scalar perturbations one gets
\begin{eqnarray} 
m\sim \sqrt{\frac{3}{10}}\pi (1-n_s)\times 10^{-4}M_{pl},
\end{eqnarray}
which for $n_s=0.968$,  leads to}
\begin{eqnarray} m=5\times 10^{-6}M_{pl} \quad  \mbox{and} \quad
\varphi_*=-15.81 M_{pl}. 
\end{eqnarray}

\

{{} Once again,} using event-driven integration with an ode RK78 integrator one gets
$\dot{\varphi}_{kin}=2.34\times 10^{-6} M_{pl}^2$, and thus
 \begin{eqnarray}
  H_{kin}=\frac{\dot{\varphi}_{kin}}{\sqrt{6} M_{pl}}\cong 9.5\times 10^{-7} M_{pl},
  \end{eqnarray}
  {{} and 
  \begin{eqnarray}
  \rho_{\varphi, kin}\cong  2.73\times 10^{-12}M_{pl}^4.  \end{eqnarray}
  }

{{}  To end this Section, note that, for the model (\ref{PVimproved}),  at the beginning of kination the energy density of the inflation is
$\rho_{\varphi, kin}\cong  6.26\times 10^{-12}M_{pl}^4$, which shows that the energy density 
drops an order of magnitude between the end of inflation and the beginning of kination, 
because
at the end of inflation {{} the effective Equation of State (EoS) parameter $w_{eff}=\frac{P}{\rho}$ is equal to $-1/3$, meaning that, at that moment,
$\dot{\varphi}^2=V(\varphi)$, i.e.  $\rho=\frac{3}{2}V$, and thus, at the end of inflation, as we have already seen, when $\epsilon=1\Longrightarrow 
e^{\sqrt{\frac{2}{3}}\frac{\varphi}{M_{pl}}}=\sqrt{3}(2-\sqrt{3}) $ the energy density is given by
$\frac{3}{2}\lambda (4-2\sqrt{3})^2 M_{pl}^4\cong 3.8\times 10^{-11} M_{pl}^4$.  Finally, note that the same happens for the model (\ref{pv}).}}

\section{Reheating via gravitational particle production}
Since the second derivative of the potential {{} (\ref{PVimproved})} is discontinuous at $\varphi=0$,  
from the conservation equation one can see that the third temporal derivative of the inflation field is discontinuous at the beginning of kination, and using the Raychaudhuri 
equation {{} $\dot{H}=-\frac{\dot{\varphi}^2}{2M_{pl}^2}$} one can deduce that
at the beginning of kination the third derivative of the Hubble parameter is discontinuous,  enhancing the particle production as discussed in \cite{kolb}.

Then, 
in order that vacuum polarization effects do not disturb the dynamics of the $\varphi$-field, the mass of 
superheavy $A$-particles, produced gravitationally, has to satisfy $m_A\gg H_B\gg m$,  where we have assumed that the beginning of inflation
occurs at GUT scales, that is, when the Hubble parameter is of the order $H_B\sim 10^{14}$ GeV (see for instance \cite{hyp}). {{} For this reason, 
the mass of superheavy particles must satisfy $m_A\geq 10^{15}$ GeV.}

\

In fact,  in the conformally coupled case,  {{} the $k$-mode satisfy the equation 
\begin{eqnarray}
\chi''_k+\omega_k^2(\tau)\chi_k=0,
\end{eqnarray}
where the derivative is with respect the conformal time and $\omega_k(\tau)=\sqrt{k^2+a^2(\tau)m_A^2}$ is the time dependent frequency.

\

{{} Note that the jump discontinuity of the third derivative of the Hubble rate is equivalent to a jump discontinuity of the fourth derivative of the frequency
$\omega_k(\tau)$, and thus its fifth temporal derivative is like a Dirac's delta, so for a smoother version of the potential (\ref{PVimproved}) the discontinuity of the second
derivative of the potential could be replaced by the no-adiabatic condition $\frac{1}{\omega_k^6(\tau)}\frac{d^5\omega(\tau)}{d\tau^5}\geq 1$ during a short period of time centered at the beginning of kination. However, a smoother potential hinders the possibility to obtain analytic expressions of the energy density of the produced particles, and for this reason we will continue with the potential (\ref{PVimproved}).

}

\

Then,
using the WKB approximation up to order two 
\begin{eqnarray}\label{WKB}
\chi_{2,k}^{WKB}(\tau)\equiv \sqrt{\frac{1}{2W_{2,k}(\tau)}}e^{-i\int^{\tau}W_{2,k}(s)ds},
\end{eqnarray}
where $W_{2,k} $ has the following complicated form  \cite{Bunch}
\begin{eqnarray}\label{W}
W_{2,k}=\omega_k-\frac{m^2a^4}{4\omega_k^3}(\dot{H}+3H^2)+\frac{5m^2a^6}{8\omega_k^5}H^2+\nonumber\\
\frac{m^2a^6}{16\omega_k^5}(\dddot{H}+15\ddot{H}H+10\dot{H}^2+86\dot{H}H^2+60H^4)-\nonumber \\
\frac{m^4a^8}{32\omega_k^7}(28\ddot{H}H+19\dot{H}^2+394\dot{H}H^2+507H^4)+\nonumber \\
\frac{221m^6a^{10}}{32\omega_k^9}(\dot{H}+3H^2)H^2-\frac{1105m^8a^{12}}{128\omega_k^{11}}H^4,
\end{eqnarray}
one can find the Bogoliubov coefficients of the $k$-mode, namely $\alpha_k$ and $\beta_k$, matching  the mode (\ref{WKB}) 
with the combination  $\alpha_k\chi_{2,k}^{WKB}(\tau)+\beta_k(\chi_{2,k}^{WKB})^*(\tau)$
at  $\varphi_{kin}=0$, i.e., when the third derivative of the Hubble rate is
discontinuous.
Denoting by $\tau_{kin}$ this time, we will have $\beta_k=i{\mathcal W}[\chi_{2,k}^{WKB}(\tau_{kin}^-),\chi_{2,k}^{WKB}(\tau_{kin}^+)]$,
where ${\mathcal W}$ is the Wronskian, and we have used the notation 
$f(\tau_{kin}^{\pm})=
\lim_{\tau\rightarrow \tau_{kin}^{\pm}}
f(\tau)$, i.e., $f(\tau^{\pm}_{kin})$ denotes the limit on the right and on the left of the point $\tau_{kin}$.

\

Then,
the leading term of the $\beta_k$-Bogoliubov coefficient is $\frac{1}{2}\left( \frac{W_{2,k}(\tau_{kin}^-)-W_{2,k}(\tau_{kin}^+)}{\sqrt{W_{2,k}(\tau_{kin}^+)W_{2,k}(\tau_{kin}^-)}}  \right)$, and  thus, form the expression (\ref{W}), one can see that the discontinuous term is 
$\frac{m^2a^6}{16\omega_k^5}\dddot{H}$, meaning that}
 {\begin{eqnarray}
 |\beta_k|^2\cong 
 \frac{m_A^4a_{kin}^{12}(\dddot{H}(\tau_{kin}^-)-\dddot{H}(\tau_{kin}^+))^2}{1024\omega^{12}_k(\tau_{kin})}.
 \end{eqnarray}}
 
 {{}
 Therefore, 
 deriving the Raychaudury equation twice, one has 
 $\dddot{H}=-\frac{1}{M_{pl}^2}(\ddot{\varphi}^2+\dot{\varphi} \dddot{\varphi})$, obtaining
 \begin{eqnarray}\label{A}
 \dddot{H}(\tau_{kin}^-)-\dddot{H}(\tau_{kin}^+)=\nonumber \\ -\frac{1}{M_{pl}^2} \dot{\varphi}_{kin}( \dddot{\varphi}(\tau_{kin}^-)-\dddot{\varphi}(\tau_{kin}^+)  ).
 \end{eqnarray}
 
 In addition, from the temporal derivative of the conservation equation, $\dddot{\varphi}+3\dot{H}\dot{\varphi}+3H\ddot{\varphi}+\dot{\varphi}V_{\varphi\varphi}=0$, and taking into account that
 $V_{\varphi\varphi}(0^+)=0$, one deduces that {{} for the model (\ref{PVimproved})
 \begin{eqnarray}\label{B}
 \dddot{\varphi}(\tau_{kin}^-)-\dddot{\varphi}(\tau_{kin}^+)=-\dot{\varphi}_{kin}V_{\varphi\varphi}(0^-)\nonumber \\ =-\frac{4}{3}\lambda\dot{\varphi}_{kin}M_{pl}^2,  \end{eqnarray}
 thus, using (\ref{A}) and (\ref{B}), one gets
 \begin{eqnarray}
 \dddot{H}(\tau_{kin}^-)-\dddot{H}(\tau_{kin}^+)=\frac{4}{3}\lambda\dot{\varphi}_{kin}^2
  \end{eqnarray}
  and the expression of the square  of the $\beta_k$-Bogoliubov coefficient becomes
   {\begin{eqnarray}
 |\beta_k|^2\cong 
 \frac{m_A^4 \lambda^2a_{kin}^{12}\dot{\varphi}_{kin}^4}{576\omega^{12}_k(\tau_{kin})}.
  \end{eqnarray}}  
  }
\

 On the other hand,
 the energy density of the produced particles
 $\rho_A(\tau)= \frac{1}{2\pi^2a^4(\tau)}\int_0^{\infty}\omega_k(\tau) k^2 |\beta_k|^2 dk\ $ (see for instance \cite{Birrell}),
 before the decay of the $X$-particles, evolves as
{{} \begin{eqnarray}\label{particleproduction}
\rho_A(\tau)\cong \frac{m_A}{2\pi^2a^3(\tau)}\int_0^{\infty} k^2 |\beta_k|^2 dk\nonumber \\
\cong 3.7\times10^{-6}\lambda^2\left( \frac{\dot{\varphi}_{kin}}{m_A}\right)^4\left( \frac{a_{kin}}{a(\tau)} \right)^3,\end{eqnarray}}
where $A=X, Y$.

\begin{remark}
Note that creation of superheavy particles in this model is power law small. Effectively,  {{}$\rho_A(\tau_{kin})\sim \left( \frac{\dot{\varphi}_{kin}}{m_A}\right)^4$} and this is due to the discontinuity
of the second derivative of the potential at $\varphi=0$. On the contrary, when the potential {{} is} very smooth the energy density of the created superheavy particles is exponentially suppressed by a factor {{}$e^{-c_Am_A/H_{kin}}$} \cite{kolb2}, where  $c_A$ is a model-dependent dimensionless parameter, meaning that for such a class of potentials the gravitational particle production mechanism is not efficient. 
\end{remark}
}

\

Thus, before the decay of the $X$-particles, one will have
\begin{eqnarray}
\rho_Y(\tau)=\left( \frac{m_X}{m_Y} \right)^4\rho_X(\tau),
\end{eqnarray}
which means that, for the PV model, one has to assume $m_X\ll m_Y$     in order to have a radiation era.

\

At this point, it is important to take into account  that when reheating is due to the gravitational production of superheavy particles, in order that the overproduction of GWs
does not alter the  BBN success,  the decay of these particles has to take place after the end of kination \cite{hyp}. Then, assuming as usual instantaneous thermalization, the reheating is produced immediately after the decay of the $X$-particles, obtaining
\begin{eqnarray}
\rho_{Y,rh}=\left( \frac{m_X}{m_Y} \right)^4\rho_{X,rh},
\end{eqnarray}
where, the subindex "rh" means that the quantities are evaluated at the reheating time.
After reheating,  the evolution of the corresponding energy densities will be
\begin{eqnarray}
\rho_X(\tau)=\rho_{X,rh}\left(\frac{a_{rh}}{a(\tau)} \right)^4,  \rho_Y(\tau)=\rho_{Y,rh}\left(\frac{a_{rh}}{a(\tau)} \right)^3,\end{eqnarray}
meaning that at the matter-radiation equality:
\begin{eqnarray}
\frac{a_{rh}}{a_{eq}}=\frac{\rho_{Y,rh}}{\rho_{X,rh}}=\left( \frac{m_X}{m_Y} \right)^4,
\end{eqnarray}
and consequently
\begin{eqnarray}
\rho_{Y,eq}=\rho_{Y,rh}\left( \frac{m_X}{m_Y} \right)^{12}
=\frac{\pi^2 g_*}{30}T_{rh}^4\left( \frac{m_X}{m_Y} \right)^{16},\end{eqnarray}
where $T_{rh}$ is the reheating temperature and $g_*=106.75$ are the degrees of freedom for the Standard Model.

\

On the other hand, 
 considering the central values obtained in \cite{planck} 
  of  the red shift at the matter-radiation equality $z_{eq}=3365$,
the present value of the ratio of the matter energy density to the critical one $\Omega_{m,0}=0.308$, and $H_0=67.81\; \mbox{Km/sec/Mpc}$,
one can deduce that  the present value of the matter energy density is $\rho_{m,0}=3H_0^2M_{pl}^2\Omega_{m,0}=3.26\times 10^{-121} M_{pl}^4$, and at matter-radiation equality one will 
have $\rho_{m,eq}=\rho_{m,0}(1+z_{eq})^3=
4.4\times 10^{-1} \mbox{eV}^4$. Since practically all the matter has a not baryonic origin, one can conclude that $\rho_{Y,eq}\cong \rho_{m,eq}$,
meaning that the reheating temperature is given by a function of $m_Y/m_X$ as follows:
{{}\begin{eqnarray}\label{temperature1}
{T}_{rh}\cong 3.3\times 10^{-10}\left( \frac{m_Y}{m_X} \right)^{4} \mbox{ GeV}.\end{eqnarray}}

\subsection{Decay after the end of the kination regime}
As we have already explained in the previous Section, in order that the overproduction of GWs  does not alter the BBN success, the decay of the $X$-particles has to be produced after the end
of kination, which occurs when the energy density of the inflaton field is equal to the one of the $X$-particles, {{} i.e.,  when $\rho_X(\tau_{end})=\rho_{\varphi}(\tau_{end})$, where we have denoted by $\tau_{end}$ the time at which kination ends. } Then,  the decaying rate, namely $\Gamma$, has to satisfy
  ${\Gamma}\leq H(\tau_{end})\equiv H_{end}$, and one has 
\begin{eqnarray}\label{31}
H^2_{end}=\frac{2\rho_{\varphi, end}}{3M_{pl}^2}, \end{eqnarray}
and 
{{} \begin{eqnarray}\rho_{\varphi, end}=\rho_{\varphi, kin}\left( \frac{a_{kin}}{a_{end}} \right)^6=3H^2_{kin}M_{pl}^2 \left( \frac{a_{kin}}{a_{end}} \right)^6.
\end{eqnarray}

Now, taking into account that during kination the energy density of the inflaton field decays as $a^{-6}$, and the one of the produced particles as $a^{-3}$, 
at the end ok kination ($\rho_{X, end}=\rho_{\varphi, end}$),  we will have 
\begin{eqnarray}
\rho_{X, kin}\left( \frac{a_{kin}}{a_{end}} \right)^3=\rho_{\varphi, kin}\left( \frac{a_{kin}}{a_{end}} \right)^6,\end{eqnarray}
that is,
$\left( \frac{a_{kin}}{a_{end}} \right)^3=
\frac{\rho_{X,kin}}{\rho_{\varphi, kin}}$, 
and introducing  the so-called {\it heating efficiency} defined in \cite{rubio} as
{{}\begin{eqnarray}\Theta\equiv 
\frac{\rho_{X,kin}}{\rho_{\varphi, kin}}\cong 7.2\times 10^{-38}\left(\frac{M_{pl}}{m_X} \right)^4,\end{eqnarray}}
we can write $\rho_{\varphi, end}=3H^2_{kin}M_{pl}^2\Theta^2$.

}

\

Consequently,  (\ref{31}) leads to  $H_{end}=\sqrt{2}H_{kin}\Theta$, and 
from the constraint $\Gamma\leq H_{end}$
one obtains the bound
{{}\begin{eqnarray}\label{const1}
\frac{\Gamma}{M_{pl}}\leq 1.5 \times 10^{-43}\left(\frac{M_{pl}}{m_X} \right)^4.\end{eqnarray}}

On the other hand,  assuming  once again instantaneous thermalization,  
the energy density of the $X$-particles at the reheating time will be $\rho_{ X,rh}=3{\Gamma}^2M_{pl}^2$, and thus,
the reheating temperature  will be
given by:
\begin{eqnarray}\label{temperature2}
T_{rh}=
\left( \frac{90}{\pi^2 g_*} \right)^{\frac{1}{4}}\sqrt{{\Gamma}M_{pl}}
\cong 1.3 \times 10^{18} \sqrt{\frac{\Gamma}{M_{pl}}} \mbox{ GeV}.
\end{eqnarray}

As a consequence, from the two expressions of the reheating temperature (\ref{temperature1}) and (\ref{temperature2}) one can write the mass of the dark matter
as a function of $\Gamma$ and $m_X$ as follows:
{{}\begin{eqnarray}\label{darkmatter}
m_Y\cong 7.9 \times 10^6\left(\frac{\Gamma}{M_{pl}} \right)^{1/8}m_X.
\end{eqnarray}}

{{}
\begin{remark}
In our work we have not considered the production of light particles nearly conformally coupled with gravity \cite{ford} because its energy never dominates and do not have any influence in the evolution of the Universe. Effectively, the energy density of these light particles, namely $\rho_r$, evolves as \cite{ford,pv} (see also \cite{haro11} for a detailed discussion)
\begin{eqnarray}\rho_r(\tau)\cong 10^{-2}(1-6\xi)^2H_{kin}^4\left(\frac{a_{kin}}{a(\tau)} \right)^4,\end{eqnarray} where $\xi$ is the coupling constant and  for the sake of simplicity we will take $|1-6\xi|\sim 10^{-2}$, although it could be smaller than  $10^{-2}$.

Then, when the energy density of the $X$-particles is of the same order than of the field $\varphi$, one has $\rho_{r,end}\cong 10^{-6}H^4_{kin}\Theta^{4/3}$ which has to
be compared with $\rho_{\varphi,end}=3H_{kin}^2M_{pl}^2\Theta^2$. A simple calculation leads to
{{}\begin{eqnarray}
\frac{\rho_{r,end}}{\rho_{\varphi,end}}\cong 1.3\times 10^{-3}\left(10^{14}\frac{m_X^4}{M_{pl}^4}\right)^{2/3},
\end{eqnarray}}
and for masses satisfying $m_X\leq 3\times 10^{-3}M_{pl}\cong 7.3\times 10^{15}$ GeV, which as we will see enter in our range, we have
{{}\begin{eqnarray}
\frac{\rho_{r,end}}{\rho_{\varphi,end}}\leq 0.53,
\end{eqnarray}}
concluding that the energy density of the light particles created during the phase transition from the end of inflation to the beginning of kination never dominates because
its energy density decreases  as $a^{-4}$ while the one of $X$-particles as $a^{-3}$.
\end{remark}
}

\subsection{Overproduction of GWs }
\label{sec-overproduction}
The success of the BBN demands that the ratio of the energy density of GWs to the one of the produced particles at the reheating time satisfies \cite{hossain3}
\begin{eqnarray}\label{bbnconstraint}
\frac{\rho_{GW, rh}}{\rho_{X,rh}}\leq 10^{-2},
\end{eqnarray} 
where the energy density of the GWs is given by $\rho_{GW}(\tau)\cong 10^{-2} H^4_{kin} \left(a_{kin}/a(\tau) \right)^4$  (see for instance \cite{ford}).

\

{{} Therefore, taking into account that
\begin{eqnarray}
\left( \frac{a_{{kin}}}{a_{end}} \right)^4=\Theta^{4/3},
\end{eqnarray}
and
\begin{eqnarray}
\left( \frac{a_{{end}}}{a_{rh}} \right)^4=\left(\frac{\rho_{X,rh}}{\rho_{X,end}}  \right)^{4/3}=\left(\frac{\Gamma}{\sqrt{2}H_{kin}\Theta}\right)^{8/3},
\end{eqnarray}
 writing $\frac{a_{{kin}}}{a_{rh}} =\left( \frac{a_{{kin}}}{a_{end}} \right)\left( \frac{a_{{end}}}{a_{rh}} \right)$}
we will have 
\begin{eqnarray}
\rho_{GW, rh}=
10^{-2} H^4_{kin}
\left( \frac{\Gamma}{\sqrt{2\Theta}H_{kin} }  \right)^{8/3},
\end{eqnarray}
and thus, 
{{}\begin{eqnarray}
\frac{\rho_{GW, rh}}{\rho_{X,rh}}\cong 7.2\times 10^{38}\left( \frac{m_X}{M_{pl}} \right)^{16/3}\left( \frac{\Gamma}{M_{pl}} \right)^{2/3},
\end{eqnarray}}
meaning that
the bound (\ref{bbnconstraint}) leads to the constraint
{{}\begin{eqnarray}\label{const2}
\frac{\Gamma}{M_{pl}}\leq 5.1 \times 10^{-62}\left(\frac{M_{pl}}{m_X} \right)^8.
\end{eqnarray}}

{
Here, it is important to realize that for 
{{}$m_X\geq 2.4 \times 10^{-5} M_{pl}$ } 
the constraint 
(\ref{const2}) automatically implies (\ref{const1}), and thus, taking into account that $T_{rh}> 1$ MeV because the BBN occurs at the MeV regime \cite{gkr}, one gets that
$\Gamma$ must satisfy 
{{}\begin{eqnarray}\label{const3} 
5.9\times 10^{-43}\leq
\frac{\Gamma}{M_{pl}}\leq 5.1 \times 10^{-62}\left(\frac{M_{pl}}{m_X} \right)^8
\end{eqnarray}}
which always holds when 
{{}\begin{eqnarray}
5.8\times 10^{13} \mbox{ GeV}\leq m_X\leq 10^{16} \mbox{ GeV}.
\end{eqnarray}}

{{} Taking into account that $m_X\geq 10^{15}$ GeV, (recall that, as we have explained in Section II,  
$m_X\gg H_B\sim 10^{14}$ GeV) the mass of $X$-particles is constrained to 
$
10^{15} \mbox{ GeV}\leq m_X\leq 10^{16} \mbox{ GeV}, 
$ and }
consequently, from (\ref{temperature2}) and (\ref{const3}), for our model  the reheating temperature is bounded by
{{}\begin{eqnarray}
1 \mbox{ MeV}\leq T_{rh}\leq 
9.7 \mbox{ GeV},
\end{eqnarray}}
and  from   (\ref{darkmatter}) and (\ref{const3}) the mass of the $Y$-particles by
{{}{}\begin{eqnarray}
4.1\times 10^{16} \mbox{ GeV}\leq m_Y\leq 4.1 \times 10^{17}\mbox{ GeV}.
\end{eqnarray}}

\

{ We finish this Section with the following remark: As we can see, the choice of masses of the $X$-field greater than $10^{15}$  GeV produce a very low reheating temperature. However, as has been discussed in the introduction of \cite{attractor1} (see also the end of the Section 4.2 in \cite{attractor0} and the bound obtained in
\cite{rubio}), when reheating is via gravitational production of light particles, for very low temperatures less than $10^4$ GeV,  a spike in the Gravitational Wave spectrum, which is large enough to challenge the BBN process,  is generated during kination. To overpass this situation we have to consider masses of the $X$-field   satisfying the condition $5.8\times 10^{13}\leq m_X< 10^{14}$ GeV, because in this situation, 
 taking   $\frac{\Gamma}{M_{pl}}= 5.1 \times 10^{-62}\left(\frac{M_{pl}}{m_X} \right)^8$ in (\ref{const3}),  one gets
 \begin{eqnarray}
T_{rh}= 2.93\times 10^{-13}\left( \frac{M_{pl}}{m_X}\right)^4 \mbox{GeV},
 \end{eqnarray}
 which for $m_X< 10^{14}$ GeV, leads to the lower bound $T_{rh}\geq 10^4$ GeV.
 
 \

 Another way to alleviate this situation is to assume that the $X$-field is not conformally coupled with gravity. In this situation,  the $X$-field could have masses of the order 
 $10^{15}$ GeV or greater, obtaining a maximum  reheating temperature of $66$ TeV (see \cite{hs19} for a detailed discussion).
 
 \
 
 Finally, as we will see in next Section, when the particles responsible for the reheating are created via instant preheating this problem disappear,  because the reheating temperature is around $10^8$ GeV.
}

\


{
\section{Instant preheating}
In this   Section 
 we consider an
interaction between the scalar field and a massless  $X$-field   conformally coupled with gravity, whose interacting Lagrangian is given by
 $ {\mathcal L}_{int}=-\frac{1}{2}g^2\varphi^2 X^2$, where $g$  is a coupling constant
 {{} and the enhanced symmetry point has been chosen $\varphi=0$, because, as we have already seen, at this point the velocity of the scalar field is nearly maximum (see Figure $2$), what, as one can see from formula (\ref{M}), maximizes the particle production.}
 In this situation $X$-particles, {{} having an effective mass
 $m_{eff}=g\varphi(t)$,}
  are created via a mechanism named {\it instant 
 preheating},  which was   introduced in \cite{fkl0}  in the framework of standard inflation, and was applied, for the first time, to quintessential inflation in \cite{fkl}.

{{}\begin{remark}
Note that here the $X$-field is completely different to the one considered in the previous Sections, however the superheavy dark matter  $Y$-field is the same, i.e., it continues only interacting gravitationally.
\end{remark}}

As was discussed in \cite{fkl}, in order to avoid a second inflationary period,  it is mandatory that,  {{} unlike the superheavy particles created gravitationally studied in the previous section,}
these $X$-particles   decay well before the end of kination. Then, at the matter-radiation equality we will have
\begin{eqnarray}
\rho_{X,eq}=\rho_{X,dec}\left( \frac{a_{dec}}{a_{eq}}  \right)^4, \quad \rho_{Y,eq}=\rho_{Y,dec}\left( \frac{a_{dec}}{a_{eq}}  \right)^3,\end{eqnarray}
and {{} since $\rho_{X,eq}=\rho_{Y,eq}$ one will have}
{{}\begin{eqnarray}
\rho_{Y,eq}=\rho_{Y, dec}\left(\frac{\rho_{Y,dec}}{\rho_{X,dec}}\right)^3.
\end{eqnarray}}

Now, using that  the decay of the $X$-particles is finished when $\Gamma=H_{dec}=H_{kin}\left( \frac{a_{kin}}{a_{dec}}\right)^3$, 
{{} and  that the energy density of the $Y$-particles decreases  as $a^{-3}$, i.e.,  $\rho_{Y,dec}=\rho_{Y,kin}\left(\frac{a_{kin}}{a_{dec}} \right)^3$,}
  we obtain
{{}\begin{eqnarray}
\rho_{Y,eq}=
\rho_{Y, kin}\frac{\Gamma}{H_{kin}} \left(\frac{\rho_{Y,dec}}{\rho_{X,dec}}\right)^3.
\end{eqnarray}

{{} In addition,}
taking into account that at the decay time the scalar field is near $M_{pl}$  (see for details \cite{fkl}), and thus, 
the effective mass of the $X$-particles is $gM_{pl}$, one gets
\begin{eqnarray}
\rho_{X,dec}=gM_{pl}n_{X,dec}=gM_{pl}n_{X,kin}\left(\frac{a_{kin}}{a_{dec}}\right)^3,
\end{eqnarray}
where $n_X$ denotes the number density of produced $X$-particles.

Therefore,    one will have
\begin{eqnarray}
\frac{\rho_{Y,dec}}{\rho_{X,dec}}=\frac{\rho_{Y,kin}}{gM_{pl}n_{X,kin}}.\end{eqnarray}

}

On the other hand, at the beginning of kination the {{} number density of  $X$-particles is \cite{fkl}
\begin{eqnarray}\label{M}
n_{X,kin}=\frac{g^{3/2}\dot{\varphi}_{kin}^{3/2}}{8\pi^3},
\end{eqnarray}
 and the energy density of the $Y$-particles} is given by the formula (\ref{particleproduction}), meaning that,
{at the matter-radiation equality one has
{{}\begin{eqnarray}
\rho_{Y,eq}=\rho_{Y,kin}\frac{\Gamma}{H_{kin}}\left(\frac{8\pi^3\rho_{Y,kin}}{g^{5/2}M_{pl}\dot{\varphi}_{kin}^{3/2}}  \right)^3\cong \nonumber\\
 9.6\times 10^{-53} g^{-15/2}\left(\frac{M_{pl}}{m_Y} \right)^{16}\frac{\Gamma}{M_{pl}}\mbox{ eV}^4,
\end{eqnarray}}
which compared with the observational 
 value of the matter density at the matter-radiation equality  $4.4\times 10^{-1} \mbox{eV}^4$, leads to
{{}\begin{eqnarray}
m_Y\cong 5.9\times 10^{-4} g^{-15/32}\left( \frac{\Gamma}{M_{pl}}\right)^{1/16} M_{pl}.
\end{eqnarray}}

}

Dealing with the reheating temperature, if one assumes once again instantaneous thermalization, it is given by (see  \cite{Haro} for details)
\begin{eqnarray}
T_{rh}=\left( \frac{30}{g_* \pi^2} \right)^{1/4}\rho_{X,dec}^{1/4}\sqrt{\frac{\rho_{X,dec}}{\rho_{\varphi, dec}}}\nonumber \\
\cong 10^{14} g^{15/8}\left(\frac{M_{pl}}{\Gamma}\right)^{1/4} \mbox{ GeV},
\end{eqnarray}
because at the end of the decay of the $X$-particles 
\begin{eqnarray}
\rho_{\varphi,dec}=3\Gamma^2M_{pl}^2 \quad \mbox{ and } \nonumber \\
\rho_{X,dec}\cong 10^{-2} g^{5/2}\sqrt{\frac{H_{kin}}{M_{pl}}}\frac{\Gamma}{M_{pl}}M_{pl}^4.
\end{eqnarray}

 {
When  $X$-particles decay
into fermions { via a Yukawa type interaction $h\psi\bar{\psi}X$} with a decaying rate
${\Gamma}=\frac{h^2 gM_{pl}}{8\pi}$, where $h$ is a coupling constant \cite{fkl}, the mass
of the $Y$-particles and 
the reheating temperature become
\begin{eqnarray}
{{} m_Y\cong 1.1\times 10^{15} g^{-13/32}h^{1/8} \mbox{ GeV}} \quad \mbox{ and } \nonumber \\
T_{rh}\cong 2.2\times 10^{14} {g^{13/8}}{{h}^{-1/2}} \mbox{ GeV}.
\end{eqnarray}

However, as has been showed in \cite{Haro} there is a narrow range of values of the parameters $g$ and $h$ for which  {\it instant preheating} is viable. For  example, 
choosing ($h=10^{-1}$, $g=10^{-4}$) or ($h=10^{-2}$, $g=5\times 10^{-5}$) one gets:
{{}\begin{eqnarray}
m_Y\sim 10^{16}
 \mbox{ GeV and  }
 T_{rh}\sim 2.2\times 10^8 \mbox{ GeV} .
\end{eqnarray}

 }
 
 \

Finally, we want to stress that when the particle production of $X$-particles is via {\it instant preheating} the overproduction of GWs does not alter the success of the BBN,
because 
\begin{eqnarray}
\frac{\rho_{GW, rh}}{\rho_{X,rh}}\leq \frac{\rho_{GW, kin}}{gM_{pl}n_{X,kin}}
\cong 1.6\times 10^{-16}g^{-5/2}\leq 10^{-5}. 
\end{eqnarray}

{{}

\section{Evolution of the universe in quintessential inflation}

This section is a review of \cite{hap18} and the Section $4$ of \cite{hs19}.

\

We start with the initial conditions, at the beginning of kination for the model (\ref{PVimproved}), obtained in Section $2$:
\begin{eqnarray}
\varphi_{kin}=0, \quad \dot{\varphi}_{kin}=3.54\times 10^{-6} M_{pl}^2.
\end{eqnarray}

During kination, the scale factor and the Hubble rate evolves as $a\propto t^{1/3}\Longrightarrow H=\frac{1}{3t}$, and from the Friedmann equation, the evolution in this phase will be
\begin{eqnarray}
\frac{\dot{\varphi}^2}{2}=\frac{M_{pl}^2}{3t^2}
\Longrightarrow 
\varphi(t)=\sqrt{\frac{2}{3}}M_{pl}\ln \left( \frac{H_{kin}}{H(t)} \right).\end{eqnarray}

Then, 
at the end of kination, one has 
 \begin{eqnarray}
\varphi_{end}=-\sqrt{\frac{2}{3}}M_{pl}\ln\left( \sqrt{2}\Theta \right), 
\dot{\varphi}_{end}=2\sqrt{3}M_{pl}H_{kin}\Theta,
\end{eqnarray}
where, once again, we have used the relation $H_{end}=\sqrt{2}H_{kin}\Theta$.

\

During the period between $t_{end}$ and $t_{rh}$, in the case that the $X$ particles are created gravitationally, the universe is matter dominated and, thus, the Hubble parameter becomes $H=\frac{2}{3t}$. During this epoch, the gradient of the potential could also be disregarded,  hence, the equation of the scalar field becomes $\ddot{\varphi}+\frac{2}{t}\dot{\varphi}=0$, and thus, after few calculations, at the reheating time one has
\begin{eqnarray}
\varphi_{rh}\cong
\varphi_{end}+\sqrt{\frac{2}{3}}M_{pl}
\end{eqnarray}
and
\begin{eqnarray}
\dot{\varphi}_{rh}
=\frac{\sqrt{3}}{4}\frac{M_{pl}H_{rh}^2}{H_{kin}\Theta}
\end{eqnarray}

During the radiation period one can continue disregarding the potential,  obtaining
\begin{eqnarray}
\varphi(t)=\varphi_{rh}+2\dot{\varphi}_{rh}t_{tr}\left(1-\sqrt{\frac{t_{rh}}{t}}\right),
\end{eqnarray}
and thus, 
since $\dot{\varphi}_{rh}t_{tr}= \frac{\pi}{6}\sqrt{\frac{ g_*}{30}}\frac{T^2_{rh}}{H_{kin}\Theta}$ (in \cite{hap18} and \cite{hs19} wrongly the authors take 
$\dot{\varphi}_{rh}t_{tr}=\sqrt{\frac{2}{3}}M_{pl}$)
 at the matter-radiation equality one has
\begin{align}\label{eq}
 \varphi_{eq} =\varphi_{rh}+     \frac{\pi}{3}\sqrt{\frac{ g_*}{30}}\frac{T^2_{rh}}{H_{kin}\Theta}\left(1-\sqrt{\frac{4H_{eq}}{3H_{rh}}}\right)\nonumber\\
 =\varphi_{rh}+ \frac{\pi}{3}\sqrt{\frac{ g_*}{30}}\frac{T^2_{rh}}{H_{kin}\Theta} 
 \left(1-\sqrt{\frac{4}{3}}\left( \frac{g_{eq}}{g_{*}} \right)^{1/4}\frac{T_{eq}}{T_{rh}}\right) \nonumber \\
\cong \varphi_{rh}+ \frac{\pi}{3}\sqrt{\frac{ g_*}{30}}\frac{T^2_{rh}}{H_{kin}\Theta}\cong \varphi_{rh}+\frac{2T^2_{rh}}{H_{kin}\Theta},
  \end{align}
 where $g_{eq}\cong 3.36$ are the degrees of freedom at the matter-radiation equality and $T_{eq}$ is the temperature of the radiation at the matter-radiation equilibrium, which is related with the energy
 density via the relation $\rho_{eq}=\frac{\pi^2}{15}g_{eq}T^4_{eq}\cong 8.8\times 10^{-1} \mbox{eV}^4$ , and thus, given by $T_{eq}\cong 7.9\times 10^{-10}$ GeV.
 
In the same way,  
 \begin{align}\label{doteq}
\dot{\varphi}_{eq}=\dot{\varphi}_{rh}\frac{t_{rh}}{t_{eq}}\sqrt{\frac{t_{rh}}{t_{eq}}}=\left(\frac{16g_{eq}}{9g_*}\right)^{3/4}\left(\frac{T_{eq}}{T_{rh}}\right)^3 \dot{\varphi}_{rh}
\nonumber\\
\cong 1.7\frac{T_{eq}^3T_{rh}}{M_{pl} H_{kin}\Theta}\cong 5.8 \times 10^{-46}\frac{T_{rh}m_X^4}{M_{pl}^3}.
\end{align}

\

After the matter-radiation equality the dynamical equations can not be solved analytically and, thus, one needs to use numerics to compute them. In order to do that, we need to use a ``time'' variable  that we choose to be minus the number of $e$-folds up to the present epoch, namely, $N\equiv -\ln(1+z)=\ln\left( \frac{a}{a_0}\right)$. Now, using  the variable $N$,  one can recast the  energy density of radiation (the energy density of the decay products of the $X$-field which we continue denoting by $\rho_X$) and dark matter respectively as
\begin{eqnarray}
 \rho_{X}(N)= \frac{\rho_{eq}}{2}e^{4(N_{eq}-N)} ,
 \rho_{Y}(N)=\frac{\rho_{eq}}{2}e^{3(N_{eq}-N)},
\end{eqnarray}
where  
$N_{eq}=-\ln(1+z_{eq})=-8.121$ is the value of $N$ at the matter-radiation equality.

In order to obtain the dynamical system for the (\ref{PVimproved}) model, we 
introduce the following dimensionless variables
 \begin{eqnarray}
 x=\frac{\varphi}{M_{pl}}, \qquad y=\frac{\dot{\varphi}}{H_0 M_{pl}},
 \end{eqnarray}
 where $H_0\cong 1.42 \times 10^{-33}$ eV denotes the current value of the Hubble parameter. Now, using the variable
 $N = - \ln (1+z)$ defined above and also using the conservation equation $\ddot{\varphi}+3H\dot{\varphi}+V_{\varphi}=0$, we will have the following  non-autonomous dynamical system:
 \begin{eqnarray}\label{system}
 \left\{ \begin{array}{ccc}
 x^\prime & =& \frac{y}{\bar H}~,\\
 y^\prime &=& -3y-\frac{\bar{V}_x}{ \bar{H}}~,\end{array}\right.
 \end{eqnarray}
 where the prime represents the derivative with respect to $N$, $\bar{H}=\frac{H}{H_0}$   and $\bar{V}=\frac{V}{H_0^2M_{pl}^2}$. Moreover, the Friedmann equation now looks as  
 \begin{eqnarray}\label{friedmann}
 \bar{H}(N)=\frac{1}{\sqrt{3}}\sqrt{ \frac{y^2}{2}+\bar{V}(x)+ \bar{\rho}_{X}(N)+\bar{\rho}_{Y}(N) }~,
 \end{eqnarray}
where we have introduced the following dimensionless energy densities
 $\bar{\rho}_{X}=\frac{\rho_{X}}{H_0^2M_{pl}^2}$ and 
 $\bar{\rho}_{Y}=\frac{\rho_{Y}}{H_0^2M_{pl}^2}$.

Then, we have to integrate the dynamical system, starting at $N_{eq}=-8.121$, with initial condition $x_{eq}$ and $y_{eq}$ which are obtained analytically in  
formulas (\ref{eq}) and (\ref{doteq}). The value of the parameter $\tilde{M}$ is obtained equaling at $N=0$ the equation (\ref{friedmann}) to $1$, i.e., imposing $\bar{H}(0)=1$.

\

On the other hand,
note that 
\begin{eqnarray}
y_{eq}=4.1\times 10^{-4} \frac{T_{rh}}{\mbox{GeV}}\left(\frac{m_X}{M_{pl}}  \right)^4,
\end{eqnarray}
and thus, for viable reheating temperatures $T_{rh}\leq 9.7$ GeV one has $y_{eq}\ll 1$.
And
for $x_{eq}$, after a simple calculation, one gets
\begin{eqnarray}
x_{eq}\cong\sqrt{\frac{2}{3}}\left(86.17-4\ln\left(\frac{M_{pl}}{m_X}\right)\right)
\nonumber \\
+1.9\times 10^{41}\frac{T_{rh}^2}{M_{pl}^2}\left(\frac{m_X}{M_{pl}}\right)^4.\end{eqnarray}

\

{

 \begin{figure}
\includegraphics[width=0.5\textwidth]{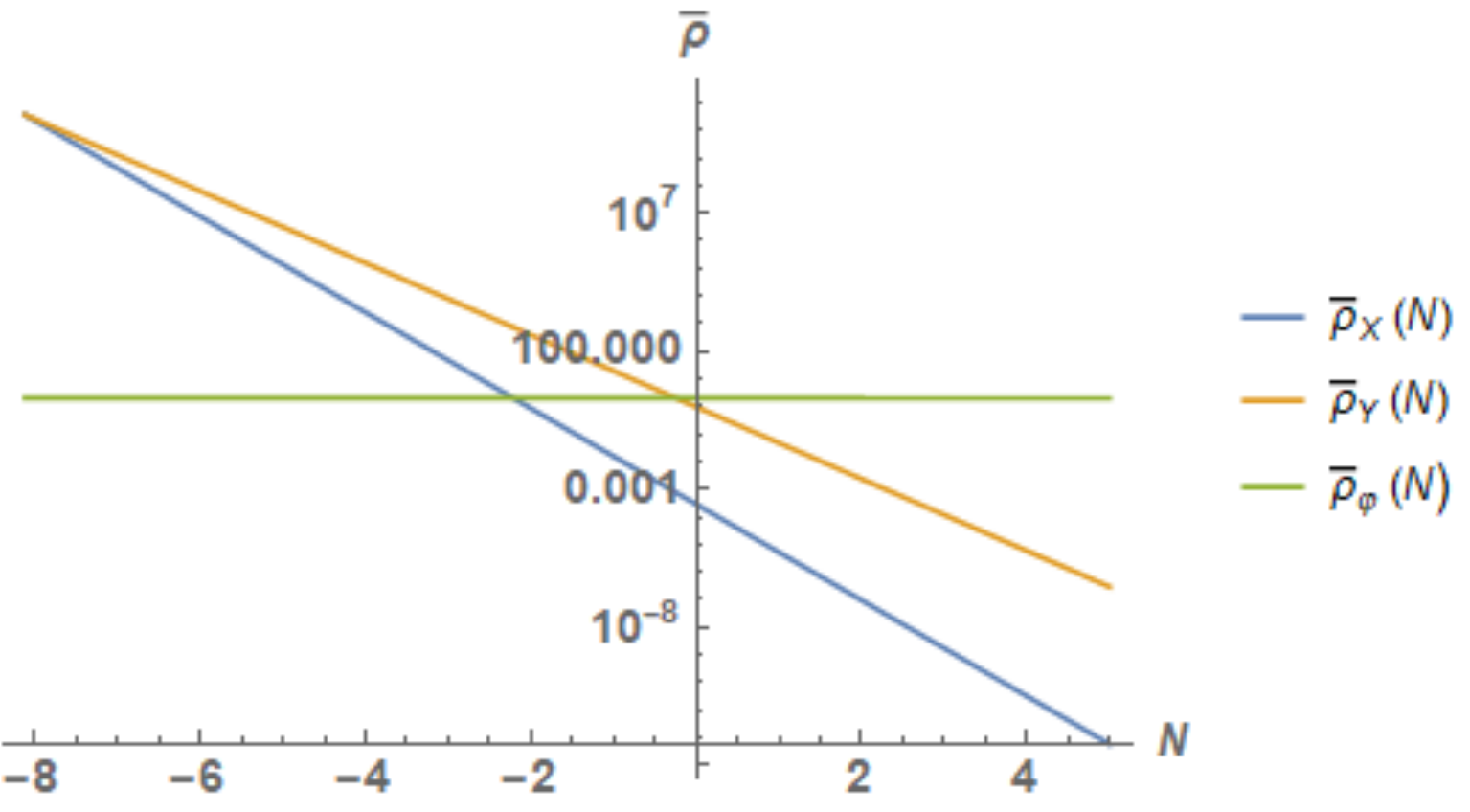}
{{}\caption{Evolution of the different dimensionless energy densities for $m_X \sim 10^{14}$ GeV and $T_{rh}\sim 10^4$ GeV.}}
\end{figure}

\begin{figure}
\includegraphics[width=0.5\textwidth]{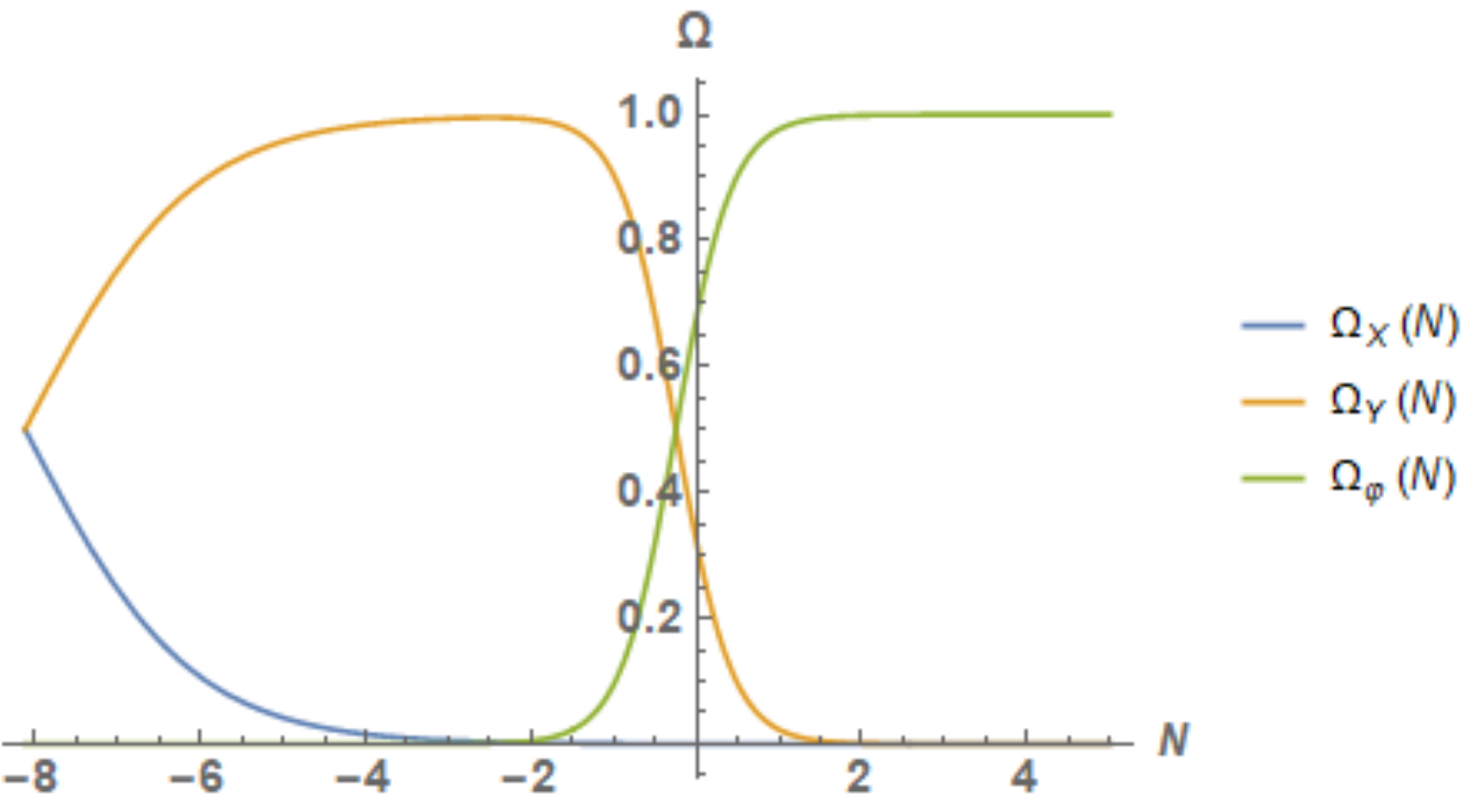}
{{}\caption{Evolution of the different $\Omega$ for $m_X\sim  10^{14}$ GeV and $T_{rh} \sim 10^4$ GeV.
At late times $\Omega_{\varphi}=1$, meaning that all the energy density of the universe is the one of the scalar field.}}
\end{figure}

\begin{figure}
\includegraphics[width=0.5\textwidth]{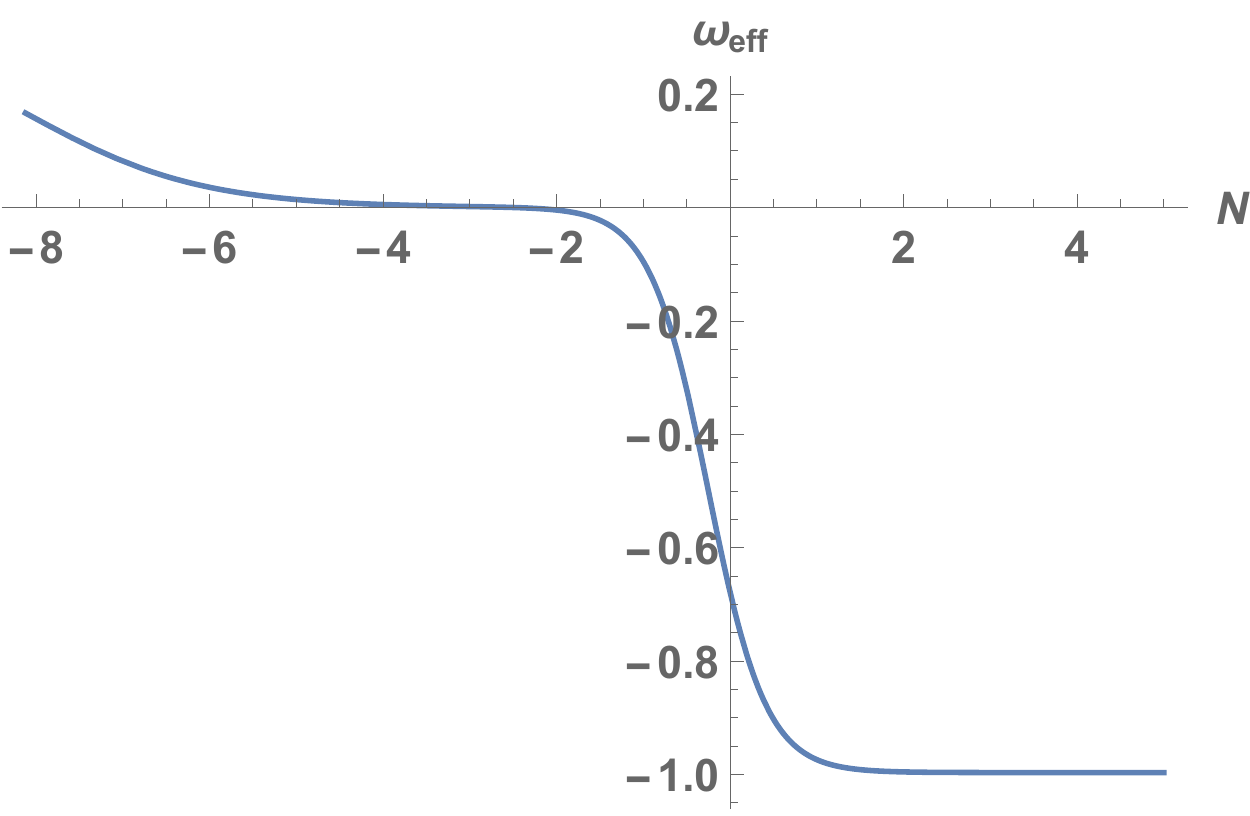}
{{}\caption{Evolution of the effective Equation of State  parameter for $m_X\sim  10^{14}$ GeV and $T_{rh}\sim  10^4$ GeV. Al late times $w_{eff}\rightarrow -1$, what means that the universe accelerates forever entering in a de Sitter phase.}}
\end{figure}

\

Summing up, 
what we have obtained numerically for the viable values of the reheating temperature and the masses of the $X$-filed is that the value of the mass $\tilde{M}$ ranges 
between $2.5\times 10^5$ and $8.6\times 10^5$ GeV, what completely agrees with the value obtained by Peebles and Vilenkin in his seminal paper \cite{pv}.
In addition, as one can see in Figure $3$ that the scalar field {{} slow-rolls the inverse power law potential} after the matter-radiation equality,  leading to 
an eternal acceleration because the effective Equation of State parameter goes towards $-1$ (see Figure $5$).

\

{{}  Finally note that the for the potential (\ref{PVimproved}) the energy scale of inflation is \cite{verde} $V^{1/4}(\varphi\ll -M_{pl})\sim \lambda^{1/4}M_{pl}
\sim 10^{15}$ GeV, which is very close to the GUT scales, while the energy scale for Dark Energy  $V^{1/4}(\varphi\cong 0)\sim \lambda^{1/4}\tilde{M}
\sim 10^{2}$ GeV is near the electroweak scale. Therefore,  our model provides   natural scales for inflation and Dark Energy.

}

\section{Conclusions}

In this paper we have presented the idea of creating  dark matter in a quintessential inflation model whose potential, which is an improvement of the well-known Peebles-Vilenkin one, is composed by a Starobinsky Inflationary type-potential matched with an inverse power law potential, which is responsible for quintessence. Since the phase transition from the end of inflation to the beginning of kination is very abrupt, the adiabatic regime is broken and superheavy particles could be gravitationally produced. We have assumed two different reheating mechanisms: 

\begin{enumerate}
\item
In the first one, two kind of superheavy particles are gravitationally produced. $X$-particles,  whose decay products form the baryonic matter, and 
GIMP
$Y$-particles, which are responsible for the dark matter abundance. For this model we have shown that, for reasonable masses of the $X$-particles between
 { $10^{14}$} and  $10^{16}$ GeV, a viable model with a reheating temperature from  the MeV  to the {TeV} regime is obtained when the mass of the dark matter particles is of the order $10^{16} - 10^{17}$ GeV. 
\item The second mechanism is the well known {\it instant preheating}, where now the $X$-field is massless and coupled with the scalar field, and the superheavy $Y$-field depicting dark matter continues only interacting gravitationally. In this situation, a viable model requires a reheating temperature around  $10^8$ GeV and dark matter particles with masses around $10^{16}$ GeV.

\end{enumerate}

\

Finally, in the case that both kind of particles are produced gravitationally, we have shown numerically that the model leads, at late times, 
 to an eternal inflation  with and Equation of State parameter equal to $-1$.

}}

\

{\it Acknowledgments:} {{} I would like to thank my colleague Llibert Arest\'e Sal\'o for his valuable help in doing the numerical calculations, {{} to my other colleague Jaume Amor\'os for reading all the previous versions of the manuscript,  and also to the referees for
your suggestions that have been essential to improve this work}.}
This investigation has been supported by MINECO (Spain) grant
 MTM2017-84214-C2-1-P, and  in part by the Catalan Government 2017-SGR-247.

\end{document}